\documentclass[onecolumn, pra, showpacs,superscriptaddress]{revtex4}
\usepackage{amssymb}
\usepackage{graphicx}
\usepackage{dcolumn}
\usepackage{bm}
\usepackage{amsmath}

\begin{document}

\title{Dynamical evolution of an effective two-level system with $\mathcal{PT%
}$ symmetry}
\author{Lei Du}
\affiliation{Institute of Theoretical Physics, Shanxi University, Taiyuan 030006, P. R.
China}
\author{Zhihao Xu}
\affiliation{Institute of Theoretical Physics, Shanxi University, Taiyuan 030006, P. R.
China}
\affiliation{Collaborative Innovation Center of Extreme Optics, Shanxi University,
Taiyuan 030006, P.R.China}
\affiliation{State Key Laboratory of Quantum Optics and Quantum Optics Devices, Institute
of Opto-Electronics, Shanxi University, Taiyuan 030006, P.R.China}
\author{Chuanhao Yin}
\affiliation{Institute of Physics, Chinese Academy of Sciences, Beijing 100080, P. R.
China}
\author{Liping Guo}
\email{guolp@sxu.edu.cn}
\affiliation{Institute of Theoretical Physics, Shanxi University, Taiyuan 030006, P. R.
China}
\affiliation{Collaborative Innovation Center of Extreme Optics, Shanxi University,
Taiyuan 030006, P.R.China}
\affiliation{State Key Laboratory of Quantum Optics and Quantum Optics Devices, Institute
of Opto-Electronics, Shanxi University, Taiyuan 030006, P.R.China}

\begin{abstract}
We investigate the dynamics of parity- and time-reversal ($\mathcal{PT}$)
symmetric two-energy-level atoms in the presence of two optical and a
radio-frequency (rf) fields. The strength and relative phase of fields can
drive the system from unbroken to broken $\mathcal{PT}$ symmetric regions.
Compared with the Hermitian model, Rabi-type oscillation is still observed,
and the oscillation characteristics are also adjusted by the strength and
relative phase in the region of unbroken $\mathcal{PT}$ symmetry. At
exception point (EP), the oscillation breaks down. To better understand the
underlying properties we study the effective Bloch dynamics and find the
emergence of the z components of the fixed points is the feature of the $%
\mathcal{PT}$\ symmetry breaking and the projections in x-y plane can be
controlled with high flexibility compared with the standard two-level system
with $\mathcal{PT}$ symmetry. It helps to study the dynamic behavior of the
complex $\mathcal{PT}$\ symmetric model.
\end{abstract}

\volumeyear{year}
\volumenumber{number}
\issuenumber{number}
\eid{identifier}
\date[Date text]{date}
\received[Received text]{date}
\revised[Revised text]{date}
\accepted[Accepted text]{date}
\published[Published text]{date}
\startpage{1}
\endpage{6}
\maketitle

In the past decades non-Hermitian Hamiltonians describing open physical
systems have attracted increasing research interests, with particular
attentions paid to a class of $\mathcal{PT}$ symmetric Hamiltonians of which
spectrum might be completely real-valued \cite{Bender}. A surge of work has
devoted to their experimental implementation in diverse physical systems,
ranging from optical waveguide structures \cite{Guo,Ruter}, flat microwave
cavities \cite{Bittner} and optical cavities \cite%
{Xiao1,Zhang,Khajavikhan,Khajavikhan2}, electronic circuits \cite{Schindler}
and whispering-gallery modes \cite{Yang1} to mesh lattices \cite{Wimmer}.
The relevant properties of $\mathcal{PT}$ symmetric Hamiltonians have been
extensively studied, such as eigenvalues, eigenfunctions and dynamical
evolution. Among all of $\mathcal{PT}$ symmetric systems, a model of two
coupled modes subjected to gain and loss with equal amplitudes is of
particular interest, because it is the minimal system to reveal $\mathcal{PT}
$ physics \cite{Yaakov,Bender2,Graefe,xiao,Panahi}. The extension of such
two modes to many modes with $\mathcal{PT}$ symmetry also exists, such as
so-called Bose-Hubbard dimer \cite{Niederle2,Niederle,Chen} and a
one-dimensional tight-binding chain with two conjugated imaginary potentials
at two edges \cite{Song}.

In experiments, a two-level atom coupled with a near resonant radiation
field can describe a coupled two-mode system. The coupling between two
levels can be also implemented in many ways. Such as two coupled hyperfine
levels can be produced by adiabatically elimination of the third
intermediate level using Raman lasers in a $\Lambda $-type three-level
atomic system \cite{shore,Alexanian,Wang1}. The two hyperfine energy levels
can be directly dressed by a microwave field or a rf field \cite{Hemmer}.
Manipulating this three-level system by the microwave field and Raman lasers
together may generate some interesting phenomena. Scully group discovered
that electromagnetically induced transparency can be controlled by the
relative phase between Raman lasers and the microwave field \cite{Scully}.
Much more physics has been involved in the presence of both couplings \cite%
{HG,Helm,Palmer}. Gain and loss can also be incorporated into
coupled-two-level atoms, realizing\textbf{\ }two-energy-level systems with $%
\mathcal{PT}$- or anti-$\mathcal{PT}$\ symmetry \cite%
{Guo,Niederle2,Niederle,yanhong,Bender3,Rabi,Wunner}. In addition, more
energy levels may be designed to satisfy $\mathcal{PT}$ symmetry by lasers
controlling, such as the interactions between lasers and a three-level or
four-level atom \cite{Konotop,Huang,Chan,Hao}.

\begin{figure}[tbp]
\includegraphics[width=3.2in]{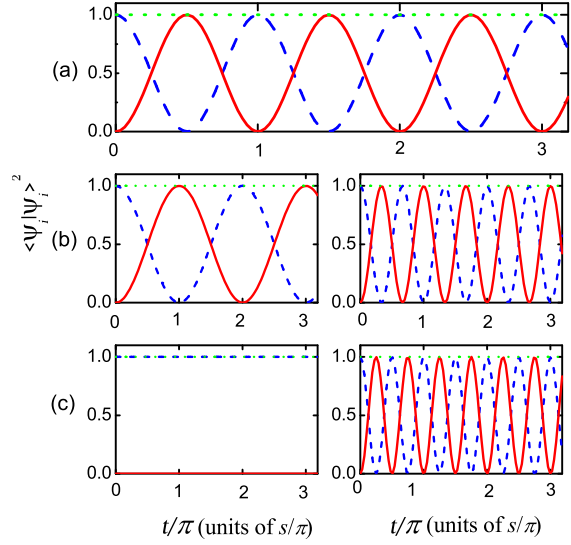}

\caption{Probability evolution of different values of $\Omega $ for $\protect%
\gamma =0$: (a) $\Omega =0$; (b) $\Omega =0.5$; (c) $\Omega =1.0$ with the
left $\protect\phi /2\protect\pi =0$ and the right $\protect\phi /2\protect%
\pi =0.5$. Red solid line describes the density of $\left\vert
3\right\rangle $; blue dash line describes the density of $\left\vert
1\right\rangle $; and green dot line \textbf{is} for the total density.}
\label{fig1}
\end{figure}

In this paper, we theoretically study the dynamics of the $\mathcal{PT}$
symmetric two-level atoms in three regimes i.e. unbroken $\mathcal{PT}$
symmetry, EP and broken $\mathcal{PT}$ symmetry. The paper is organized as
following, we firstly study the time evolution of Hermitian two-level system
and then we discuss eigenvalues and investigate dynamics of the effective
model with balanced gain and loss. We mainly focus on the effects of Rabi
frequency and relative phase. A summary is given. In Supplemental Material
the effectively conserved two-level Hamiltonian is derived from the $\Lambda
$-type model coupled with two optical fields and one rf field by means of
adiabatically eliminating of the third energy level.

\section{Model of the Two-Level System with $\mathcal{PT}$ symmetry}

We consider an effective two-level model with balanced gain and loss. The
effective Hamiltonian can be derived from a $\Lambda $-type three-level
system via adiabatic elimination of an excited state $\left\vert
2\right\rangle $ under the large detuning condition (see Supplemental
Material). The Hamiltonian is described as following:

\begin{align}
H_{eff}& =-i\gamma \left\vert 1\right\rangle \left\langle 1\right\vert
+\left( 1-\Omega e^{i\phi }\right) \left\vert 1\right\rangle \left\langle
3\right\vert  \notag \\
& +i\gamma \left\vert 3\right\rangle \left\langle 3\right\vert +\left(
1-\Omega e^{-i\phi }\right) \left\vert 3\right\rangle \left\langle
1\right\vert ,  \label{H}
\end{align}%
where $\left( 1-\Omega e^{\pm i\phi }\right) $ describes the effective
coupling between states $\left\vert 1\right\rangle $ and $\left\vert
3\right\rangle $, $\Omega $ is the effective strength of a rf field and
atom, $\phi $ is the relative phase between rf and two laser fields, and $%
\gamma $ is the amplitude of gain and loss.
\begin{figure}[tbp]
\includegraphics[width=3.4in]{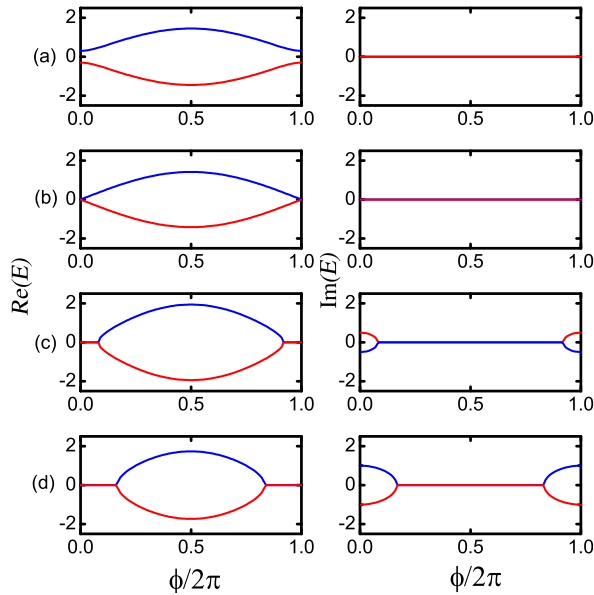}
\caption{Real (left) and imaginary (right) parts of the eigenvalues of the $%
\mathcal{PT}$-symmetric two-level system with function of the relative phase
$\protect\phi /2\protect\pi $: (a) $\Omega =0.5$, $\protect\gamma =0.4$; (b)
$\Omega =0.5$, $\protect\gamma =0.5$; (c) $\Omega =1.0$, $\protect\gamma %
=0.5 $; (d) $\Omega =1.0$, $\protect\gamma =1.0$.}
\label{fig2}
\end{figure}
In the absence of $\gamma $, the system is a standard hermitian case. The
energy spectrum of the case is two branches, \textit{i.e.} $E_{\pm }=\pm
\sqrt{\left( 1-2\Omega \cos \phi +\Omega ^{2}\right) }$. For $\Omega <1$,
the two branches are well separated by the energy gap. When $\Omega =1$, two
branches touch at $\phi =0$. When $\Omega $ is larger than $1$, the energy
gap opens again. The dynamics of the Hamiltonian (\ref{H}) can be obtained
by solving the corresponding Schr\"{o}dinger equation
\begin{equation}
i\partial _{t}\psi =H_{eff}\psi .  \label{Sch}
\end{equation}%
Considering the initial condition, $\psi \left( t=0\right) =\psi _{0}$, the
dynamic evolution equation can be integrated formally,%
\begin{equation*}
\psi \left( t\right) =\hat{U}\left( t\right) \psi _{0},
\end{equation*}%
where the time evolution operator is defined by $\hat{U}\left( t\right)
=\exp \left( -iH_{eff}t\right) $. It can be analytically obtained,
\begin{equation*}
\hat{U}\left( t\right) =\left(
\begin{array}{cc}
\cos \left( \omega _{1}t\right) & -i\frac{\left( 1-\Omega e^{i\phi }\right)
}{\omega _{1}}\sin \left( \omega _{1}t\right) \\
-i\frac{\left( 1-\Omega e^{-i\phi }\right) }{\omega _{1}}\sin \left( \omega
_{1}t\right) & \cos \left( \omega _{1}t\right)%
\end{array}%
\right) ,
\end{equation*}%
where $\omega _{1}=\left\vert 1-\Omega e^{i\phi }\right\vert $. As a
concrete example, we set the initial state as $\psi _{0}=\left( 1,0\right)
^{T}$. We can obtain the dynamics of both states at time $t$

\begin{align}
\left\vert \psi _{1}\left( t\right) \right\vert ^{2}& =\frac{1+\cos \left(
2\omega _{1}t\right) }{2},  \label{Herp} \\
\left\vert \psi _{3}\left( t\right) \right\vert ^{2}& =\frac{1-\cos \left(
2\omega _{1}t\right) }{2},  \notag
\end{align}%
where $\left\vert \psi _{1}\left( t\right) \right\vert ^{2}$ $\left(
\left\vert \psi _{3}\left( t\right) \right\vert ^{2}\right) $ is the
occupation of $\left\vert 1\right\rangle $ $\left( \left\vert 3\right\rangle
\right) $. Fig. \ref{fig1} shows the probability evolution\textbf{s} of two
states for different $\Omega $ following Rabi oscillation with the
periodicity $\pi /\omega _{1}$. For $\Omega =0$, $\omega _{1}=1$, only the
two-photon term determines the Rabi oscillation as shown in Fig. \ref{fig1}%
(a). With the increase of $\Omega $ and the relative phase $\phi =0$ (the
left column), the periodicity is larger than Fig. \ref{fig1}(a) case. When $%
\Omega =1$ and $\phi =0$, the Rabi oscillation breaks down. The coexistence
of rf-field and two optical fields coherently destructs the oscillation [the
left panel of Fig. \ref{fig1}(c)]. However, when $\phi =\pi $ the
oscillation periodicity decreases with the increase of $\Omega $ as shown in
the right column of Fig. \ref{fig1}(b)-(c). For the hermitian case, we can
see the total occupation is always unit and time-independent.

In the present of gain and loss ($\gamma \neq 0$) the eigenvalues of the $%
\mathcal{PT}$ symmetric Hamiltonian (\ref{H}) are%
\begin{equation}
E_{\pm }=\pm \sqrt{\left\vert 1-\Omega e^{i\phi }\right\vert ^{2}-\gamma ^{2}%
}.  \label{EV}
\end{equation}%
Fig. \ref{fig2} describes the real and imaginary parts of the eigenvalues of
the $\mathcal{PT}$-symmetric two-level system with function of the relative
phase $\phi $ for different $\Omega $ and $\gamma $. As shown in Fig. \ref%
{fig2}(b), when $\gamma =\left\vert 1-\Omega e^{i\phi }\right\vert $, $%
E_{+}=E_{-}$ and the eigenvectors coincide, which is defined as the EP $%
\gamma _{_{PT}}$. When $\gamma <\gamma _{_{PT}}$, the system has a purely
real spectrum and satisfies $\mathcal{PT}$ symmetry [Fig. \ref{fig2}(a)]%
\textit{.} Whereas when $\gamma >\gamma _{_{PT}}$, the eigenvalues are
complex and the symmetry is spontaneously broken [Fig. \ref{fig2}(c)-(d)].

\begin{figure}[tbp]
\includegraphics[width=3.4in]{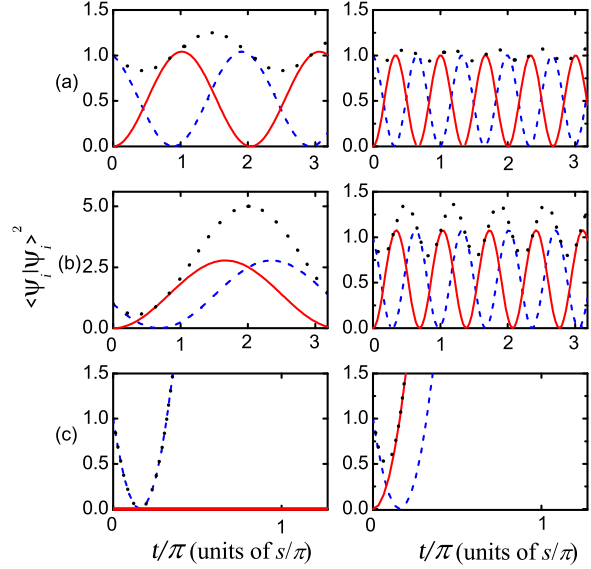}
\caption{The norm evolutions for Non-Hermitian systems:\ $\protect\phi =0$\
(left) and $\protect\pi $\ (right); The first two rows with $\Omega =0.5\ $%
and two different values of $\protect\gamma $: (a) $\protect\gamma =0.1$\
and (b) $\protect\gamma =0.4$; The last row (c) $\Omega =1.0$, $\protect%
\gamma =2.0$. Red solid line describes the norm of $\left\vert
3\right\rangle $, blue dash line is for the norm of $\left\vert
1\right\rangle $, and black dot line is for the total norm.}
\label{fig3}
\end{figure}

The dynamics of the non-Hermitian two-level quantum system can easily be
calculated from the dynamic evolution equation (\ref{Sch}). For the
Hamiltonian (\ref{H}) outside the EP, one obtains the evolution operator:%
\begin{equation}
\hat{U}\left( t\right) =\left(
\begin{array}{cc}
\cos \left( \omega t\right) -\frac{\gamma }{\omega }\sin \left( \omega
t\right) & -i\frac{\left( 1-\Omega e^{i\phi }\right) }{\omega }\sin \left(
\omega t\right) \\
-i\frac{\left( 1-\Omega e^{-i\phi }\right) }{\omega }\sin \left( \omega
t\right) & \cos \left( \omega t\right) +\frac{\gamma }{\omega }\sin \left(
\omega t\right)%
\end{array}%
\right) ,  \label{TiE}
\end{equation}%
with the angular frequency%
\begin{equation}
\omega =\sqrt{\left\vert 1-\Omega e^{i\phi }\right\vert ^{2}-\gamma ^{2}}.
\label{fre}
\end{equation}%
When\textbf{\ }$\gamma =0$,\textbf{\ }$\omega =\omega _{1}$. While, EPs ($%
\gamma =\gamma _{_{PT}}$) are singularities of the spectrum at which
spontaneous $\mathcal{PT}$ symmetry breaking has been observed in
experiments \cite{Ruter}. It can lead to some new effects, e.g., chiral
behavior and state-exchange process \cite{Rabi,Wunner}. Hence, we also focus
on the evolution of the system with the parameters approaching EP, i.e., $%
\omega \rightarrow 0$. In this limit, the time evolution operator is%
\begin{equation}
\hat{U}_{EP}\left( t\right) \approx \left(
\begin{array}{cc}
1-\gamma t & -i\left( 1-\Omega e^{i\phi }\right) t \\
-i\left( 1-\Omega e^{-i\phi }\right) t & 1+\gamma t%
\end{array}%
\right) .  \label{EP}
\end{equation}%
For example, for an initial state $\psi _{0}=\left( 1,0\right) ^{T}$, we can
find outside the EP the norms of two levels are%
\begin{align}
\left\vert \psi _{1}\left( t\right) \right\vert ^{2}& =\left\vert \cos
\left( \omega t\right) -\gamma \frac{\sin \left( \omega t\right) }{\omega }%
\right\vert ^{2},  \label{psi} \\
\left\vert \psi _{3}\left( t\right) \right\vert ^{2}& =\left\vert i\left(
1-\Omega e^{-i\phi }\right) \frac{\sin \left( \omega t\right) }{\omega }%
\right\vert ^{3},  \notag
\end{align}%
where $\left\vert \psi _{1,3}\left( t\right) \right\vert ^{2}$ are
time-dependent periodic Rabi-type oscillation functions. The period $\pi
/\omega $ is dependent on $\gamma $, $\Omega $ and $\phi $, just as the
depiction of Eq. (\ref{fre}). In detail Fig. \ref{fig3} describes some
typical cases for the initial state in level $\left\vert 1\right\rangle $.
The left column depicts the dynamics for $\phi =0$ and the right column for $%
\phi =\pi $\textbf{.} Simultaneously, the first two rows, Fig. \ref{fig3}(a)
and (b), describe the dynamics of (\ref{H}) with the same values of $\Omega $%
\ and two different values of $\gamma $ in the $\mathcal{PT}$-symmetric
region ($\gamma <\gamma _{_{PT}}$), respectively. We can find the envelope
amplitudes increase with the increasing of $\gamma $\ in column direction.
In addition, the corresponding periods of $\phi =0$\ are larger than that of
$\phi =\pi $. The last row, Fig. \ref{fig3}(c) gives two cases when $\gamma
\geqslant \gamma _{_{PT}}$. When $\Omega =1$, $\phi =0$ and $\gamma >\gamma
_{_{PT}}$ (here $\gamma _{_{PT}}=0$) due to the coexistence of rf-field and
two optical fields coherently destructs only $\left\vert \psi _{1}\left(
t\right) \right\vert ^{2}$ first decreases and then increases to infinity,
and meanwhile, $\left\vert \psi _{3}\left( t\right) \right\vert ^{2}$ is
zero all the time as shown in the left panel of Fig. \ref{fig3}(c). In fact
in the light of (\ref{psi}) when $\Omega =1$\ and $\phi =0$ $\left\vert \psi
_{3}\left( t\right) \right\vert ^{2}$\ is always zero whether $\gamma $ is
zero or not. When $\gamma =\gamma _{_{PT}}$ the wave function evolutes
according to the operator (\ref{EP}) and the corresponding norms become
\begin{align}
\left\vert \psi _{1}\left( t\right) \right\vert ^{2}& =\left\vert 1-\gamma
t\right\vert ^{2},  \label{EPpsi} \\
\left\vert \psi _{3}\left( t\right) \right\vert ^{2}& =\left\vert -i\left(
1-\Omega e^{-i\phi }\right) t\right\vert ^{2},  \notag
\end{align}%
where $\left\vert \psi _{1}\left( t\right) \right\vert ^{2}$ first decreases
and then increases, and meanwhile, $\left\vert \psi _{3}\left( t\right)
\right\vert ^{2}$ increases for ever as shown in the right panel of Fig. \ref%
{fig3}(c).\textbf{\ }In summary, in $\mathcal{PT}$ symmetric region, the
time evolution of $\left\vert \psi _{1,3}\left( t\right) \right\vert ^{2}$
is periodic Rabi-type oscillation and in $\mathcal{PT}$ symmetric broken
region the oscillations break down and instead we observe an algebraic
growth of the norms.

In order to compare with the Hermitian case, we normalize the total norm.
The renormalized state vector \cite{Niederle,Niederle2} can be defined as%
\begin{equation}
\varphi _{j}=\frac{\psi _{j}}{\sqrt{\left\vert \psi _{1}\right\vert
^{2}+\left\vert \psi _{3}\right\vert ^{2}}},  \label{Norm}
\end{equation}%
\begin{figure}[tbp]
\includegraphics[width=3.4in]{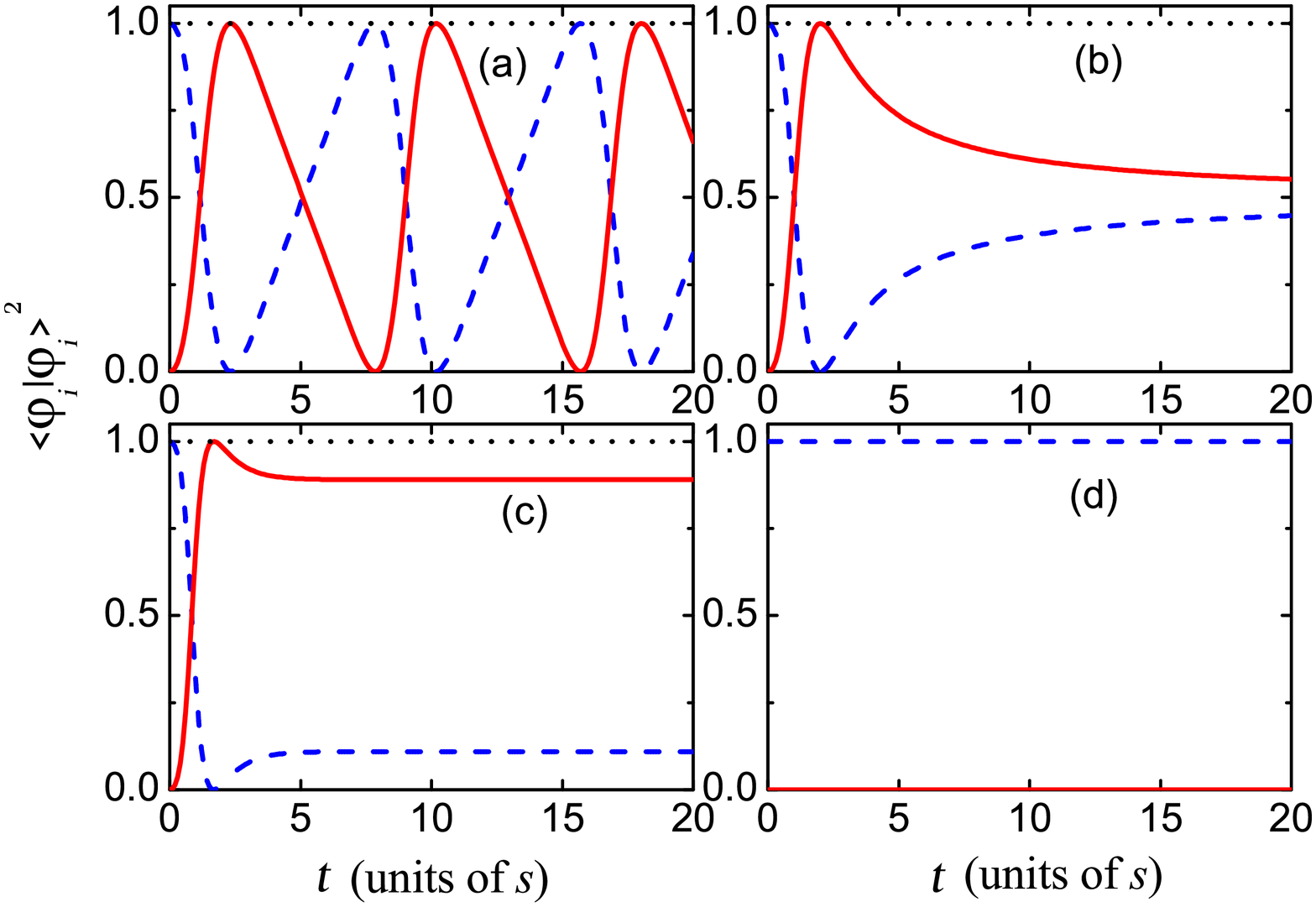}
\caption{The time evolutions of renormalized states for initial state being $%
\left( 1,0\right) ^{T}$ with $\Omega =0.5$ and different values of $\protect%
\gamma $: (a) $\protect\gamma =0.3$; (b) $\protect\gamma =0.5$; (c) $\protect%
\gamma =0.8$; And (d) $\protect\gamma =0.1$ and $\Omega =1$. Here $\protect%
\phi =0$. Red solid line describes for $\protect\varphi _{3}$, blue dash
line describes the norm of $\protect\varphi _{1}$, and black dot line is for
the time-independent $\left\vert \protect\varphi _{1}\right\vert
^{2}+\left\vert \protect\varphi _{3}\right\vert ^{2}$.}
\label{fig4}
\end{figure}
which satisfies $\left\vert \varphi _{1}\right\vert ^{2}+\left\vert \varphi
_{3}\right\vert ^{2}=1$ at any time. Then we start to deal with the time
evolution of $\varphi _{j}$ for initial state being $\left( 1,0\right) ^{T}$%
. $\left\vert \varphi _{1,3}\left( t\right) \right\vert ^{2}$\ are Rabi-type
oscillation in $\mathcal{PT}$\ symmetric region [Fig. \ref{fig4}(a)]. The
evolution tends towards stability\ in $\mathcal{PT}$\ symmetry broken regime
as shown in Figs. \ref{fig4}(b) and \ref{fig4}(c). Fig.\ \ref{fig4}(d)
describes the dynamics similar to the left of Fig. \ref{fig3}(c)\textbf{.}
To better understand the underlying properties we consider the non-Hermitian
effective Schr\"{o}dinger equation according to the Hamiltonian (\ref{H}):%
\begin{equation}
i\frac{d}{dt}\left(
\begin{array}{c}
\varphi _{1} \\
\varphi _{3}%
\end{array}%
\right) =\left(
\begin{array}{cc}
-i\gamma \left( 1-\kappa \right) & \left( 1-\Omega e^{i\phi }\right) \\
\left( 1-\Omega e^{-i\phi }\right) & +i\gamma \left( 1+\kappa \right)%
\end{array}%
\right) \left(
\begin{array}{c}
\varphi _{1} \\
\varphi _{3}%
\end{array}%
\right)  \label{SC}
\end{equation}%
where $\kappa =\left\vert \varphi _{1}\right\vert ^{2}-\left\vert \varphi
_{3}\right\vert ^{2}$. And further in the familiar way they are translated
as the Bloch vectors:
\begin{align*}
s_{x}& =\frac{1}{2}\left( \varphi _{1}^{\ast }\varphi _{3}+\varphi
_{1}\varphi _{3}^{\ast }\right) , \\
s_{y}& =\frac{1}{2i}\left( \varphi _{1}^{\ast }\varphi _{3}-\varphi
_{1}\varphi _{3}^{\ast }\right) , \\
s_{z}& =\frac{1}{2}\left( \varphi _{1}^{\ast }\varphi _{1}-\varphi
_{3}^{\ast }\varphi _{3}\right) ,
\end{align*}%
which are always restricted on the surface of the Bloch sphere and satisfy $%
s_{x}^{2}+s_{y}^{2}+s_{z}^{2}=1/4$ \cite{Niederle}. Then we can derive the
generalized Bloch equations of motion from Eq. (\ref{SC}):
\begin{figure}[tbp]
\includegraphics[width=3.0in]{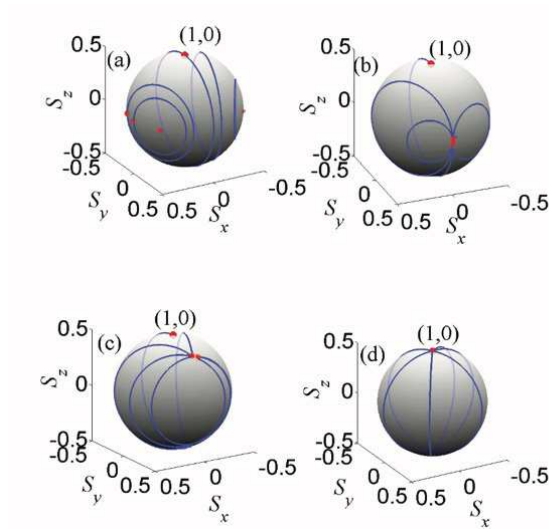}
\caption{Effective Bloch dynamics with the same parameters as in Fig. 4. Red
asterisks (*)\ embody the initial state vectors of every evolution.}
\label{fig5}
\end{figure}
\begin{align}
\dot{s}_{x}& =4\gamma s_{z}s_{x}+2\Omega s_{z}\sin \phi ,  \notag \\
\dot{s}_{y}& =4\gamma s_{z}s_{y}-2\left( 1-\Omega \cos \phi \right) s_{z},
\notag \\
\dot{s}_{z}& =-\gamma \left( 1-4s_{z}^{2}\right) +2s_{y}-2\Omega \left(
s_{y}\cos \phi +s_{x}\sin \phi \right) .  \label{motion}
\end{align}%
Thus we can get the effective Bloch dynamical evolution of this
non-Hermitian system with different parameters by the fourth-order
Runge-Kutta method. Fig. \ref{fig5} shows some typical examples of the
effective Bloch dynamics for different values of $\gamma $, $\Omega $ and $%
\phi $. When $\gamma <\gamma _{_{PT}}$, Fig. \ref{fig5}(a) shows the
evolution of arbitrary initial state is the closed circle on the surface of
Bloch sphere, which surrounds either of the two fixed points: (i)%
\begin{align}
s_{x}^{0}& =\frac{-\Omega \gamma \sin \phi \pm \omega \left( 1-\Omega \cos
\phi \right) }{2\left( 1-2\Omega \cos \phi +\Omega ^{2}\right) },
\label{PTS} \\
s_{y}^{0}& =\frac{\gamma \left( 1-\Omega \cos \phi \right) \pm \omega \Omega
\sin \phi }{2\left( 1-2\Omega \cos \phi +\Omega ^{2}\right) },  \notag \\
s_{z}^{0}& =0.  \notag
\end{align}%
The fixed points are located in x-y plane. Comparing with the standard model
in \cite{Niederle}, we have more free parameters which is helpful to control
dynamics. For example, with the change of $\Omega $ and $\phi $, $s_{y}^{0}$
are not only positive, but also can take negative values. Even in the
Hermitian case $\gamma =0$, $s_{y}^{0}$ can take non-zero values due to the
phase $\phi $ and $\Omega $. However, at EP, $\gamma =\gamma _{_{PT}}$, all
initial states would evolute along the corresponding curves and turn back
the only one fixed point: (ii)%
\begin{align}
s_{x}^{0}& =-\frac{\Omega \sin \phi }{2\gamma },  \label{PEP} \\
s_{y}^{0}& =\frac{\left( 1-\Omega \cos \phi \right) }{2\gamma },  \notag \\
s_{z}^{0}& =0,  \notag
\end{align}%
which is still located in x-y plane as shown in Fig. \ref{fig5}(b). It is
different from the only result $s_{x}^{0}=0$ mentioned in \cite{Niederle}.
Fig. \ref{fig5}(b) shows the effective Bloch dynamics for $\phi =0$ or $\pi $
where the fixed point $s_{x}^{0}=s_{z}^{0}=0$ and $s_{y}^{0}=1/2$. This
helps to investigate the potential character at EP. When $\mathcal{PT}$
symmetry breaks ($\gamma >\gamma _{_{PT}}$) there are two fixed points
again: (iii)%
\begin{align}
s_{x}^{0}& =-\frac{\Omega \sin \phi }{2\gamma },  \label{PTB} \\
s_{y}^{0}& =\frac{\left( 1-\Omega \cos \phi \right) }{2\gamma },  \notag \\
s_{z}^{0}& =\pm \frac{\left\vert \omega \right\vert }{2\gamma }.  \notag
\end{align}%
The formulas of $s_{x}^{0}$\ and $s_{y}^{0}$\ are same as (\ref{PEP}),
however, they decrease with increasing $\gamma $. Compared with (\ref{PTS})
and (\ref{PEP}) the z axis projections are no longer zero which are labeled
as '$\pm $'. Fig. \ref{fig5}(c) shows any initial states on the Bloch sphere
evolutes into one fixed point ($-$), even if initial points are near the
other one ($+$). Thus the fixed point ($+$) is defined as a source of the
dynamics and the other ($-$) is a sink of the dynamics. Fig. \ref{fig5}(d)
describes the dynamics of coherent subtraction, $\Omega =1$ and $\phi =0$,
where the fixed points are $s_{x}^{0}=s_{y}^{0}=0$ and $s_{z}^{0}=\pm 1/2$
(correspond to the states: $\left( 1,0\right) ^{T}$ and $\left( 0,1\right)
^{T}$). This is the reason that the evolution does not change with time
(since the initial state mentioned above is $\left( 1,0\right) ^{T}$) as
shown in the left panel of Figs. \ref{fig3}(c) and \ref{fig4}(d). In
addition, the emergence of $s_{z}^{0}$ is the feature of the $\mathcal{PT}$
symmetry breaking. The fixed points are important for the dynamical
evolution of $\mathcal{PT}$ symmetry systems.

\section{Summary}

In summary, we consider the effective two-level Hamiltonian with balanced
gain and loss. When the system is conserved and Hermitian, the dynamical
evolution performs Rabi oscillation and the oscillation periodicity can be
controlled by the rf frequency and relative phase. There is one exception at
which rf field balance out the actions of two optical fields and oscillation
is destructed. For the case of the effective two-level system with $\mathcal{%
PT}$ symmetry, we study the dynamics in both regions of unbroken and broken $%
\mathcal{PT}$ symmetry and at EP. Further more, using the renormalized state
vector the effective Bloch dynamical evolution has been studied in three
conditions. Since the dynamics is closely related to the fixed points, we
calculated the solutions under three parameter conditions and found the
symmetry breaks when the z component of the fixed points is not zero,
meanwhile the projection in x-y plane can be fully adjusted by the phase and
strength of optical and rf fields. It shall be helpful to the
implementation\ of related experiment.

This work is supported by the NSF of China under Grants No.11104171,
11404199, 11574187 and the Youth Science Foundation of Shanxi Province of
China under Grant No. 2012021003-1. Z. Xu is supported by the NSF of Chian
under Grants No. 11604188, NSF for youths of Shanxi Province No.
201601D201027 and 1331KSC.

\end{document}


\title{Supplemental Material: Dynamical evolution of an effective two-level
system with $\mathcal{PT}$ symmetry}
\author{Lei Du}
\affiliation{Institute of Theoretical Physics, Shanxi University, Taiyuan 030006, P. R.
China}
\author{Zhihao Xu}
\affiliation{Institute of Theoretical Physics, Shanxi University, Taiyuan 030006, P. R.
China}
\affiliation{Collaborative Innovation Center of Extreme Optics, Shanxi University,
Taiyuan 030006, P.R.China}
\affiliation{State Key Laboratory of Quantum Optics and Quantum Optics Devices, Institute
of Opto-Electronics, Shanxi University, Taiyuan 030006, P.R.China}
\author{Chuanhao Yin}
\affiliation{Institute of Physics, Chinese Academy of Sciences, Beijing 100080, P. R.
China}
\author{Liping Guo}
\email{guolp@sxu.edu.cn}
\affiliation{Institute of Theoretical Physics, Shanxi University, Taiyuan 030006, P. R.
China}
\affiliation{Collaborative Innovation Center of Extreme Optics, Shanxi University,
Taiyuan 030006, P.R.China}
\affiliation{State Key Laboratory of Quantum Optics and Quantum Optics Devices, Institute
of Opto-Electronics, Shanxi University, Taiyuan 030006, P.R.China}
\maketitle

\section{The effective Two-Level Hamiltonian}

\begin{figure}[tbp]
\includegraphics[width=3.4in]{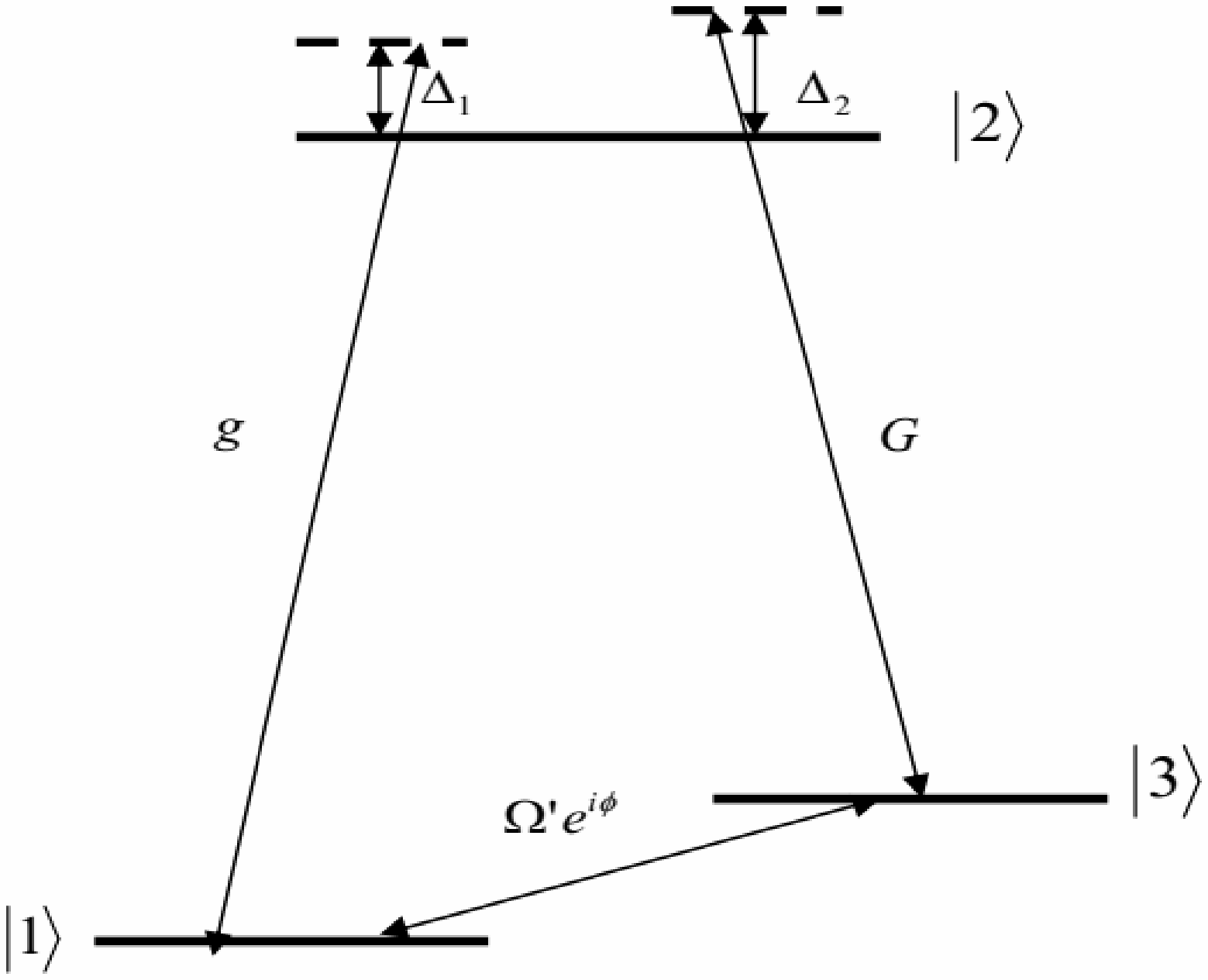}
\caption{Schematic illustration of three-level atomic system.}
\label{fig6}
\end{figure}

To get the effective two-level Hamiltonian, we start from a $\Lambda $-type
scheme shown in Fig. \ref{fig6}. Two hyperfine energy levels $\left\vert
1\right\rangle $ and $\left\vert 3\right\rangle $ are coupled by a rf-field.
Simultaneously, two optical fields couple the hyperfine levels to an excited
state $\left\vert 2\right\rangle $ \cite{Hemmer,Scully,HG,Helm}. In the
rotating frame of the two optical fields, the Hamiltonian is%
\begin{align}
H^{\prime }& =\hbar \Delta _{1}\left\vert 1\right\rangle \left\langle
1\right\vert +\hbar \Delta _{2}\left\vert 3\right\rangle \left\langle
3\right\vert  \label{HO} \\
& -\hbar \left( g\left\vert 1\right\rangle \left\langle 2\right\vert
+G\left\vert 3\right\rangle \left\langle 2\right\vert +\Omega ^{\prime
}e^{i\phi }\left\vert 1\right\rangle \left\langle 3\right\vert +H.c.\right) ,
\notag
\end{align}%
where $\Delta _{\alpha }$ ($\alpha =1,2$) are the detunings of the optical
fields from the corresponding atomic resonances, $g$ and $G$ are Rabi
frequencies of the two optical fields. $\Omega ^{\prime }$ is the coupling
strength due to the rf-field, the phase $\phi $ can be controlled via the
adjustment of the relative phase between the rf-field and two optical fields
\cite{Hemmer}. The dynamics of these three levels obeys the following
Schrodinger equation,
\begin{equation*}
i\hbar \partial _{t}\left\vert \psi \left( t\right) \right\rangle =H^{\prime
}\left\vert \psi \left( t\right) \right\rangle ,
\end{equation*}%
where the state $\left\vert \psi \left( t\right) \right\rangle $ is the
superposition of three level states,%
\begin{equation*}
\left\vert \psi \left( t\right) \right\rangle =C_{1}\left( t\right)
\left\vert 1\right\rangle +C_{2}\left( t\right) \left\vert 2\right\rangle
+C_{3}\left( t\right) \left\vert 3\right\rangle ,
\end{equation*}%
with the coefficients $C_{\alpha }\left( t\right) \equiv \left\langle \psi
|\alpha \right\rangle $ ($\alpha =1,2,3$). The equations of the coefficients
are%
\begin{align}
i\partial _{t}C_{1}& =\left( \Delta _{1}-\Delta \right) C_{1}-gC_{2}-\Omega
^{\prime }e^{i\phi }C_{3},  \notag \\
i\partial _{t}C_{2}& =-\Delta C_{2}-gC_{1}-GC_{3},  \notag \\
i\partial _{t}C_{3}& =-\Omega ^{\prime }e^{-i\phi }C_{1}-GC_{2}+\left(
\Delta _{2}-\Delta \right) C_{3},  \label{Coe}
\end{align}%
where all energies have already been translated by $-\hbar \Delta $, with $%
\Delta \equiv \left( \Delta _{1}+\Delta _{2}\right) /2$. Since $\left\vert
\Delta \right\vert \gg \left\vert \Delta _{2}-\Delta _{1}\right\vert $ and $%
\left\vert \Delta \right\vert \gg \Gamma ^{\prime }$, where $\Gamma ^{\prime
}$ is the decay rate of $\left\vert 2\right\rangle $, we can adiabatically
eliminate $C_{2}$ from the dynamics of our system, $\partial _{t}C_{2}=0$
\cite{Steck},
\begin{equation}
C_{2}=-\frac{g}{\Delta }C_{1}-\frac{G}{\Delta }C_{3}.  \label{CC}
\end{equation}%
Substituting the equation (\ref{CC}) into (\ref{Coe}), we obtain two coupled
equations for the hyperfine states,%
\begin{align*}
i\partial _{t}C_{1}& =\left( \Delta _{1}+\frac{g^{2}}{\Delta }\right)
C_{1}+\left( \frac{gG}{\Delta }-\Omega ^{\prime }e^{i\phi }\right) C_{3}, \\
i\partial _{t}C_{3}& =\left( \frac{gG}{\Delta }-\Omega ^{\prime }e^{-i\phi
}\right) C_{1}+\left( \Delta _{2}+\frac{G^{2}}{\Delta }\right) C_{3}.
\end{align*}%
Hence, the effective Hamiltonian of the above equations is%
\begin{align}
H_{eff}& =\hbar \left( \Delta _{1}+\frac{g^{2}}{\Delta }\right) \left\vert
1\right\rangle \left\langle 1\right\vert +\hbar \left( \Delta _{2}+\frac{%
G^{2}}{\Delta }\right) \left\vert 3\right\rangle \left\langle 3\right\vert
\notag \\
& +\hbar \left( \frac{gG}{\Delta }-\Omega ^{\prime }e^{i\phi }\right)
\left\vert 1\right\rangle \left\langle 3\right\vert +\hbar \left( \frac{gG}{%
\Delta }-\Omega ^{\prime }e^{-i\phi }\right) \left\vert 3\right\rangle
\left\langle 1\right\vert .  \label{H3}
\end{align}%
For simplicity, we set $g=G$, $\Delta _{1}=\Delta _{2}=\Delta $. The
effective Hamiltonian can be shown as follows

\begin{equation*}
H_{eff}=\left(
\begin{array}{cc}
0 & 1-\Omega e^{i\phi } \\
1-\Omega e^{-i\phi } & 0%
\end{array}%
\right) ,
\end{equation*}%
where the constant terms are neglected. Here $\hbar G^{2}/\Delta =1$, then $%
\Omega =\Omega ^{\prime }\Delta /G^{2}$. The related experiments for Na \cite%
{Hemmer} and Rb \cite{Scully} atomic gases have been realized. In addition,
the balanced loss or gain in our model can be realized by the control of
decays into some one auxiliary state or vice versa \cite{Chan}.